\documentclass[useAMS,usenatbib]{mn2e}
\usepackage{natbib}
\usepackage{amsmath}
\usepackage{graphicx}
\usepackage{color}
\title[Formation of a supermassive star]
{Formation of an embryonic supermassive star in the first galaxy}
\author[K. Inayoshi, K. Omukai and E. Tasker]{Kohei Inayoshi$^{1,2}$
\thanks{E-mail: inayoshi@astro.columbia.edu}, 
Kazuyuki Omukai$^{3}$
and Elizabeth Tasker$^{4}$, 
\\
$^{1}$Department of Physics, Graduate School of Science, Kyoto University, Kyoto 606-8502, Japan \\
$^{2}$Department of Astronomy, Columbia University, 550 West 120th Street, New York, NY 10027, USA\\
$^{3}$Astronomical Institute, Tohoku University, Sendai 980-8578, Japan \\
$^{4}$Department of Physics, Faculty of Science, Hokkaido University, Sapporo 060-0810, Japan}

\newcommand{\msunyr}{{\rm M}_{\sun}~{\rm yr}^{-1}}
\newcommand{\mdot}{\dot{M}_{\ast}}
\newcommand{\msun}{{\rm M}_{\sun}}

\begin{document}

\maketitle

\begin{abstract}
We studied the gravitational collapse of a warm ($\sim 8000$ K)
primordial-gas cloud as a candidate progenitor for a supermassive star
(SMS; $\ga 10^5~\msun$) using a three-dimensional hydrodynamical simulation
including all the relevant cooling processes of both H$_2$ and H,
which can potentially induce cloud fragmentation.
This is the first simulation of this kind to resolve protostar formation.
We find that 
from a weakly turbulent initial condition,
the cloud undergoes runaway 
collapse without a major episode of fragmentation. 
Although the H$_2$ fraction jumps by a large factor via the three-body reaction at $\sim 10^{-13}$ g cm$^{-3}$, 
its cooling remains inefficient due to the optical thickness,
and the temperature remains $\ga 3000$ K. 
When the central core of the cloud becomes opaque to continuum radiation
at $\sim 10^{-8}$ g cm$^{-3}$, a hydrostatic protostar 
with $\simeq 0.2~\msun$ is formed. 
The protostar grows to the mass $\simeq 1~\msun$ 
and the radius $\simeq 2$ AU within $\sim 1$ yr 
via rapid accretion of dense filamentary flows. 
With high accretion rate, $\sim 2~\msunyr$, the protostar is 
expected to turn into an SMS within its lifetime,
eventually collapsing to a seed for the supermassive black hole 
observed in the early Universe at $z\sim7$.
\end{abstract}


\section{Introduction}

Recent observations reveal the existence 
of supermassive black holes (SMBHs) with masses $\ga 10^9~\msun$ 
as early as redshift $z\ga 7$ 
\citep[e.g.,][]{2006NewAR..50..665F, 2011Natur.474..616M}.
Their very existence puts a strong constraint on the origin and 
formation pathway of SMBHs, 
since the time required to form such massive objects from stellar-mass 
seed BHs ($\sim 100~\msun$) exceeds the then Hubble time.  
As a solution to this conundrum, the formation from massive seed BHs 
originating in the collapse of supermassive stars (SMSs; $\ga 10^5~\msun$) 
has been suggested \citep[e.g.,][]{2006MNRAS.370..289B}.
Seed BHs formed in this way are expected to grow to $\ga 10^9~\msun$ by $z\sim 7$ 
via subsequent gas accretion \citep{2012ApJ...745L..29D}.

SMSs can be formed within primordial-gas clouds in massive 
haloes with virial temperature $\ga 10^4$ K, providing that 
H$_2$ cooling is prohibited throughout the protostellar collapse. 
Without this latter constraint, the gas would rapidly cool via H$_2$ 
and fragment into smaller pieces. 
H$_2$ has to therefore be dissociated e.g., via the photodissociation 
by far-ultraviolet (FUV) radiation from nearby star-forming galaxies 
\citep{O01, BL03, S10, 2011MNRAS.416.2748I}
or the collisional dissociation in the shocked gas
\citep{2012MNRAS.422.2539I}.
This in place, the cloud can collapse almost isothermally at $\sim 8000$ K 
solely by the H-atomic cooling. Previous numerical studies 
\citep[e.g.,][]{BL03, 2008ApJ...682..745W, S10} 
have implied that such a cloud collapses monolithically 
without efficient fragmentation. 
After a protostar is formed at the center, it then grows rapidly to an SMS 
via accretion from the envelope at a rate $\ga 0.1~\msunyr$. 
Under such a high accretion rate, its growth is not hindered either 
by strong radiative feedback \citep{HOY12,2013ApJ...778..178H} or 
by mass-loss due to stellar pulsations \citep{2013MNRAS.431.3036I}.

So far, however, most studies \citep[e.g.,][]{2009MNRAS.396..343R, 
2013MNRAS.433.1607L, 2013ApJ...774..149C} have utilized several 
simplifying assumptions in studying the fragmentation process, e.g., 
turning off the H$_2$ cooling, adopting optically-thin treatment of Ly$\alpha$ cooling, 
or using insufficient chemical networks that neglect the H$_2$ formation. 
Yet the efficiency of fragmentation depends strongly on the thermal evolution 
determined by the cooling processes and chemical reactions. 
As a result, it still remains unresolved whether the cloud fragments 
during the isothermal collapse at $\sim 8000$ K before forming the protostar.
In this {\it Letter}, we use a three-dimensional hydrodynamical simulation 
to study the gravitational collapse of a turbulent primordial-gas cloud with mass 
$\ga 10^5~\msun$. 
We follow the evolution until the formation of the protostar, 
including all relevant processes required for examining 
possible fragmentation during this collapse. 
Among these, the H$_2$-line cooling is of primary importance due to 
its ability to rapidly drop the temperature via thermal instability 
if the gravitational collapse is delayed, a process possible due to turbulence 
generated during the virialization of the halo. 
If the thermal instability occurs, the cloud can fragment into many smaller 
mass clumps instead of forming a single SMS. 
We therefore simulate the collapse to determine the likelihood of 
the outcome being a monolithic collapse to a single star or fragmentation 
into a binary or multiple member system.

\section{Methodology}

We performed a three-dimensional hydrodynamical simulation of 
the gravitational collapse of a primordial-gas cloud using the adaptive 
mesh refinement code, {\tt ENZO} \citep{Enzo}.
Our main purpose is to investigate the gas dynamics over a wide range of the densities
($10^{-21}\la \rho \la 10^{-7}$ g cm$^{-3}$).
The cloud initially has a spherically symmetric density profile 
enhanced by a factor $f~(=1.6)$ above the critical Bonnor-Ebert (BE) distribution,
an isothermal sphere 
embedded in a pressurized medium and
supported in marginal hydrostatic equilibrium against gravitational collapse.
According to cosmological simulations \citep[e.g.,][]{2008ApJ...682..745W}, 
at the center of a first galaxy with virial temperature $\ga 10^4$ K, 
forming in an environment where the H$_2$ formation is suppressed,
a warm ($T\sim 8000$ K) cloud with $\sim 10^5~\msun$ becomes gravitationally 
unstable at $\rho \sim 10^{-20}$ g cm$^{-3}$ and collapses.
Based on this, we set the central density and temperature of the cloud
to $\rho_c=1.67\times 10^{-20}$ g cm$^{-3}$ and $T=8000$ K, 
giving a mass and radius of $1.17\times 10^5~\msun$ and $10.8$ pc, respectively.
Although we here do not impose an external FUV radiation,  
H$_2$ is collisionally dissociated
for $\rho \ga 10^{-20}$ g cm$^{-3}$ and $T \ga 6000$K .
Note that we neglect the dark-matter gravity 
since the cloud is already bound by the self-gravity of its gas.
Our simulation box size is (50 pc)$^3$ and refinement is controlled by 
insisting that one Jeans length is resolved by at least 64 grid cells \citep[e.g.,][]{T12}. 
Under this condition, the simulation uses 23 out of the allowed 25 refinement levels, 
ensuring we are resolved by the above criteria at all times and giving a limiting 
resolution of $\la 0.1$ AU.

The development of turbulence in the central region of 
forming first galaxies has been suggested by numerical simulations
\citep[e.g.,][]{2007ApJ...665..899W,2008MNRAS.387.1021G}.
In the initial phase of collapse with $\sim 10^{-20}$ g cm$^{-3}$, 
the turbulence is still subsonic in the cloud.
To consider the density and velocity perturbations due to the turbulence,
we initially impose a subsonic velocity field (the root mean square of the velocity
is set to $0.1~c_s$) 
with power spectrum $P(k)\propto k^{-4}$, 
which corresponds to the so-called Larson's law for the contemporary star-forming regions 
\citep{1981MNRAS.194..809L}. 
To ensure that the turbulence is adequately resolved, 
we select the maximum k-mode value of 1/10 of 
the number of cells across the cloud.

\begin{figure}
\includegraphics[height=59mm,width=75mm]{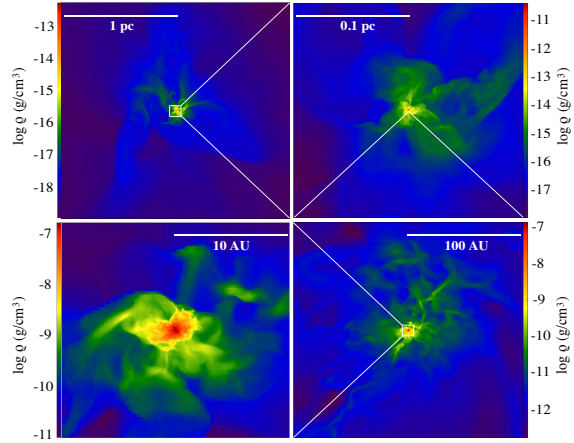}
\caption{Density distribution in the plane through the density peak
for four spatial scales:
from top-left, clockwise: the large-scale gas distribution ($\sim 1$ pc), 
a collapsing core by the H$^-$ free-bound continuum cooling ($\sim 0.1$ pc), 
the central region around the protostar ($\sim 100$ AU), and 
the final protostar ($\sim 10$ AU). 
}
\label{fig:density_large}
\end{figure}

We consider the non-equilibrium primordial chemistry of 9 species 
(H, H$_2$, e$^-$, H$^+$, H$_2^+$, H$^-$, He, He$^+$, and He$^{++}$) and 
13 hydrogen reactions selected to reproduce the correct thermal/chemical evolution
of the warm atomic-cooling cloud
\citep[reactions 3, 4, 7$-$10, 12, 15$-$18, 28, 
and 32 in table 2 of]
[]{O01}.
We adopt the reaction rate coefficients updated by the following studies:
7$-$10 \citep{2011ApJS..193....7C},
15 \citep{1996ApJ...461..265M},
17 \citep{1999ApJ...513L.147S},
and 28 \citep{1992ApJ...387...95F}.
The four helium reactions originally included in {\tt ENZO} 
are also present, although they are not relevant in our calculation. 
We initially assume a uniform distribution of
ionization degree with $10^{-4}$ and H$_2$ molecular fraction with 
$10^{-7}$, respectively \citep[e.g.,][]{S10}.
At high density, the chemical reactions proceed faster than 
the cloud collapse and chemical equilibrium is achieved.
To smoothly connect the non-equilibrium chemistry to that of equilibrium,
we solve the chemical network including both the forward and reverse
reactions for dominant processes. 
To solve the chemistry equations, we employ the piecewise exact 
solution method \citep{2008ApJ...687..303I} instead of the original {\tt ENZO} solver,
which cannot follow the chemical evolution with high enough density to reach the chemical equilibrium.
For the radiative cooling, we consider atomic cooling 
(H Ly$\alpha$, two photon emission, and H$^-$ free-bound, free-free emission)
and H$_2$ cooling (rovibrational line and collision-induced emission).
We also include the suppression of the cooling rate in the optically
thick case by using the optical depth estimated as 
$\rho \kappa L_{\rm c}$ \citep[e.g.,][]{O01, S10}, 
where $\kappa$ includes the H$_2$-line opacity and 
the Rosseland mean opacity considering the H Rayleigh scattering, 
the H$_2$ collision-induced absorption, 
and the H$^-$ bound-free and free-free absorption,
and $L_{\rm c}$ the size of the central core, which is approximately given by 
the Jeans length for the spherically symmetric cloud
in the runaway collapse.
Finally, note that we do not include the heating/cooling associated with the chemical reactions 
because their effect is negligible during the thermal evolution of the atomic-cooling clouds.

\section{Results}

\begin{figure}
\begin{center}
\includegraphics[height=120mm,width=67mm]{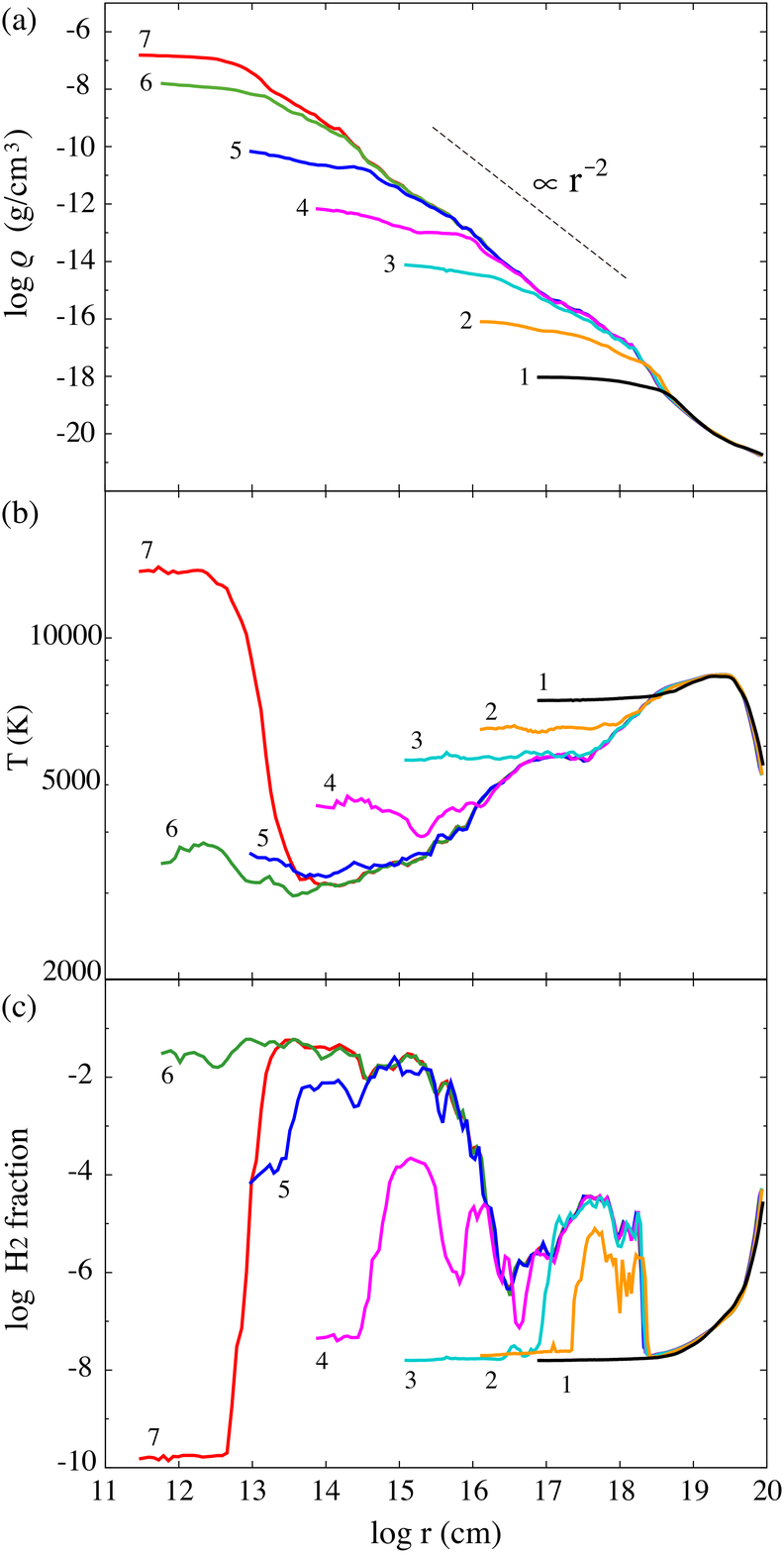}
\caption{Mass-weighted radial profiles at different evolutionary stages
of (a) mass density, (b) temperature, and (c) H$_2$ fraction.
The time sequences are indicated by numbers:
(1) $8.0\times 10^5$ yr after the initial state of our simulation.
(2) $9.3\times 10^4$ yr after (1):
the main coolants are the Ly$\alpha$ and two-photon emissions.
(3) $1.6\times 10^4$ yr after (2):
dominant cooling process shifts to the H$^-$ free-bound emission 
in the central core.
(4) $1.8\times 10^3$ yr after (3):
H$_2$ formation via the three-body reaction becomes active 
at the centre.
(5) $2.8\times 10^2$ yr after (4):
the cloud becomes optically thick to dominant H$_2$ lines.
(6) $1.8\times 10^1$ yr after (5):
the cloud becomes optically thick to the continuum opacities and 
a hydrostatic core is formed at the centre.
(7) $1.2$ yr after (6): the final state of the simulation.
}
\label{fig:pro}
\end{center}
\end{figure}

\begin{figure}
\begin{center}
\includegraphics[height=93mm,width=82mm]{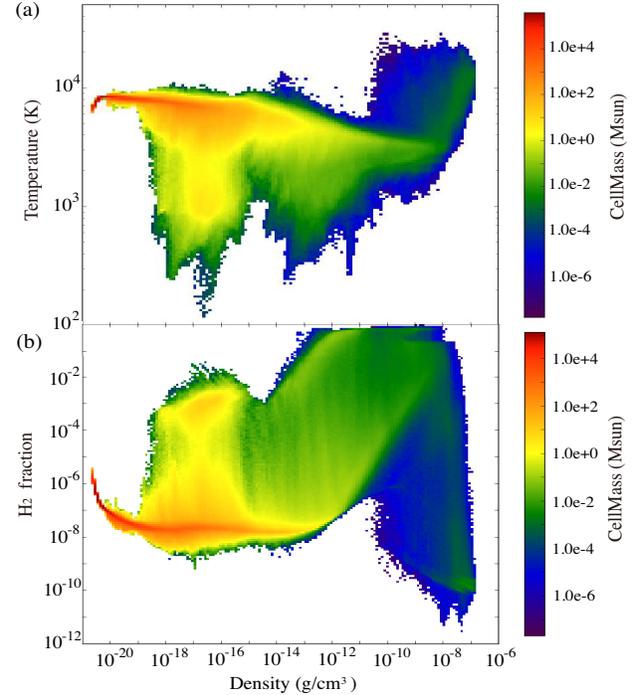}
\caption{Phase diagrams showing the distribution of (a) density-temperature and 
(b) density-H$_2$ fraction of the collapsing cloud 
at the end of the simulation.
The colors represent the total mass at the respective density and temperature.}
\label{fig:phase}
\end{center}
\end{figure}

Fig.~\ref{fig:density_large} shows the density distribution at the end of the simulation, 
where the central density reaches $\sim 10^{-7}$ g cm$^{-3}$, 
for four different spatial scales; from the top-left clockwise, 
large-scale gas distribution ($\sim 1$ pc), 
the collapsing core ($\sim 0.1$ pc), 
the central $\sim 100$ AU region, and 
the protostar formed at the center ($\sim 10$ AU). 
The central portion of the cloud undergoes the runaway collapse.
The turbulence forms filamentary structures that channel material into the central region 
($\rho \sim 10^{-8}$ g cm$^{-3}$), 
feeding the protostar.
The left-bottom panel presents the density distribution around the protostar.
At the end of this simulation, the protostellar mass reaches $\simeq 1~\msun$
and its radius $\simeq 2$ AU.
These values are consistent with the result of the stellar-structure calculation by \cite{HOY12}, 
who assumed a steady and spherical accretion.

Fig.~\ref{fig:pro} shows the evolution of mass-weighted radial profiles 
of (a) density, (b) temperature, and (c) H$_2$ fraction.
During collapse, the density profile obeys 
the self-similar solution, which consists of the central core with
flat density distribution and envelope with the $\rho \propto r^{-2}$ law 
\citep[e.g.,][]{1969MNRAS.145..271L}.
The central core collapses almost isothermally until $\sim 10^{-8}$ g cm$^{-3}$
keeping the temperature at $\sim 5000$ K.
In the low density regime of $\rho<10^{-16}$ g cm$^{-3}$, 
the cooling is mainly via the H Ly$\alpha$ and two-photon emission.
At higher density, 
the dominant cooling process shifts to the H$^-$ free-bound emission 
(H + e$^- \rightarrow$ H$^-$ + $\gamma$).
For $\ga 10^{-9}$ g cm$^{-3}$, 
photons from the H$^-$ free-bound emission are self-absorbed, as well as
Rayleigh scattered by H.
The gas cools further by the H$^-$ free-free emission 
(H + e$^- \rightarrow$ H + e$^-$ + $\gamma$)
until $\ga 10^{-8}$ g cm$^{-3}$.
Finally, the cloud becomes opaque to all those continuum opacities at 
$\sim 10^{-8}$ g cm$^{-3}$and 
a hydrostatic core, i.e., a protostar, with its mass $\simeq 0.2~\msun$
and radius $\simeq 1$ AU 
is formed at the center, where the core temperature is $\sim 4000$ K.
As the protostar grows to $\sim 1~\msun$, the temperature inside the protostar 
adiabatically increases to $\sim 10^4$ K.

The mass-weighted H$_2$ fraction initially approaches 
the equilibrium value ($\sim 10^{-8}$), where the formation through 
the electron-catalyzed reaction (H + e$^-\rightarrow$ H$^- +\gamma$; 
H$^-$ + H $\rightarrow $H$_2$ + H) and 
the collisional dissociation (H$_2$ + H $\rightarrow$ 3H) are balanced.
At $\rho \ga 10^{-13}$ g cm$^{-3}$, 
the H$_2$ fraction jumps up to $\sim 0.1$ by the three-body reaction 
(3H $\rightarrow $ H$_2$ + H) in the inner region ($r \la 10^{3}$ AU).
However, neither the H$_2$ line nor collision-induced-emission (CIE) 
cooling plays a significant role in
the thermal evolution:
H$_2$ lines are optically thick for $\rho \ga 10^{-10}$ g cm$^{-3}$ 
and other continuum cooling is more important than the H$_2$ CIE cooling.
After the protostar formation, the H$_2$ is dissociated inside owing to 
the high temperature.

Fig.~\ref{fig:phase} presents the phase diagrams showing the distribution 
of (a) temperature and (b) H$_2$ fraction as a function of the density
at the end of the simulation.
The cloud consists of two thermal phases of the gas, i.e., 
hot ($\sim $ several $10^3$ K) and 
cold ($\sim 10^3$ K) components.
As seen in Figs.~\ref{fig:pro}(b) and \ref{fig:phase}(a), 
most of the collapsing gas resides 
in the hot component, which ultimately forms a protostar at the center.
Note that since the density profile follows the self-similar form 
during the runaway collapse and 
thus the radial position has one-to-one correspondence with the density
(Fig.~\ref{fig:pro}a),
the density-temperature distribution of the hot component 
is just a reflection of the temperature profile (Fig.~\ref{fig:pro}b).
The H$_2$ fraction in the hot component remains almost constant at $\sim 10^{-8}$ 
up to $10^{-13}$ g cm$^{-3}$ and then increases almost proportionally to the density
for higher density by the three-body reaction until finally dissociated 
at $\ga 10^{-8}$ g cm$^{-3}$ as a result of the protostar formation.

Meanwhile, the cold component exists over the wide density range, 
$10^{-18}\la \rho \la 10^{-11}$ g cm$^{-3}$.
This gas is produced by the thermal instability induced by the combination
of the adiabatic cooling due to the turbulent expansion and the subsequent H$_2$ cooling.
Once the temperature decreases via adiabatic cooling, 
the H$_2$ dissociation becomes inefficient, enhancing 
the H$_2$ fraction and its cooling rate, causing the temperature to plummet. 
This process is known as the chemo-thermal instability 
associated with the H$_2$ formation/dissociation 
\citep{1979PASJ...31..505Y, 1983MNRAS.205..705S}.

\begin{figure}
\begin{center}
\includegraphics[height=39mm,width=86mm]{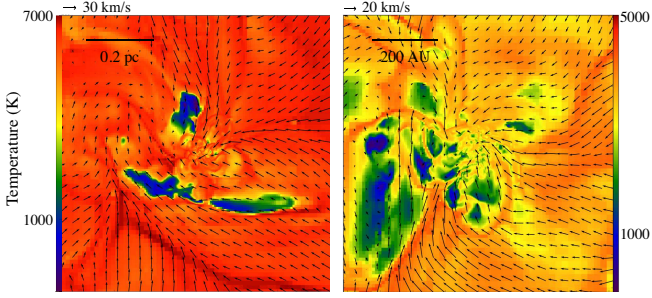}
\caption{Temperature (color) and velocity field (arrows)
in the plane through the density peak for two spatial scales. 
Cold regions are formed by the chemo-thermal 
instabilities due to the H$_2$ formation by the electron-catalyzed reaction (left)
and by the three-body reaction (right).
}
\label{fig:temp_large}
\end{center}
\end{figure}

Fig.~\ref{fig:temp_large} presents the temperature distribution and
velocity fields for two different scales.
In both panels, the coexistence of the cold and hot components is clearly visible.
Turbulence establishes a complex structure of interacting shocks and stagnation points.
The cold components in the two scales are produced 
by the thermal instabilities due to the H$_2$ formation
through the electron-catalyzed reaction ($\sim 0.2$ pc)
and three-body reaction ($\sim 200$ AU), see also Fig.~\ref{fig:phase}(b).
The cold components are not massive enough to be gravitationally bound
and have no influence on the evolution of the central collapsing region.

\begin{figure}
\begin{center}
\includegraphics[height=46mm,width=68mm]{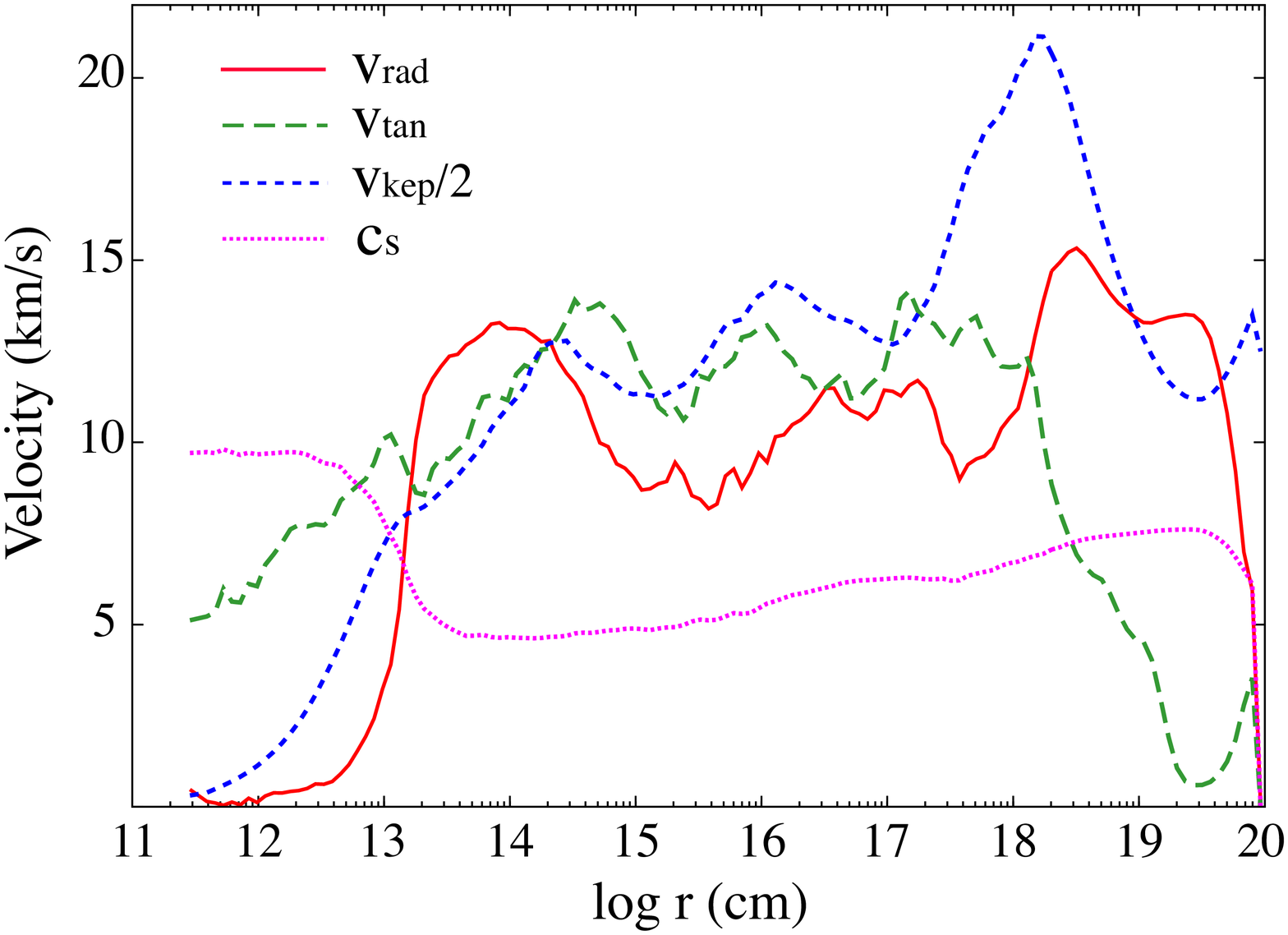}
\end{center}
\caption{
Profiles of the radial (solid) and tangential velocities (long-dashed) 
at the end of the simulation.
For comparison, a half of the Keplerian velocity (short-dashed)
and sound speed (dotted) are also shown.
All the quantities are spherically averaged. 
}
\label{fig:r_vel}
\end{figure}

\begin{figure}
\begin{center}
\includegraphics[height=46mm,width=67mm]{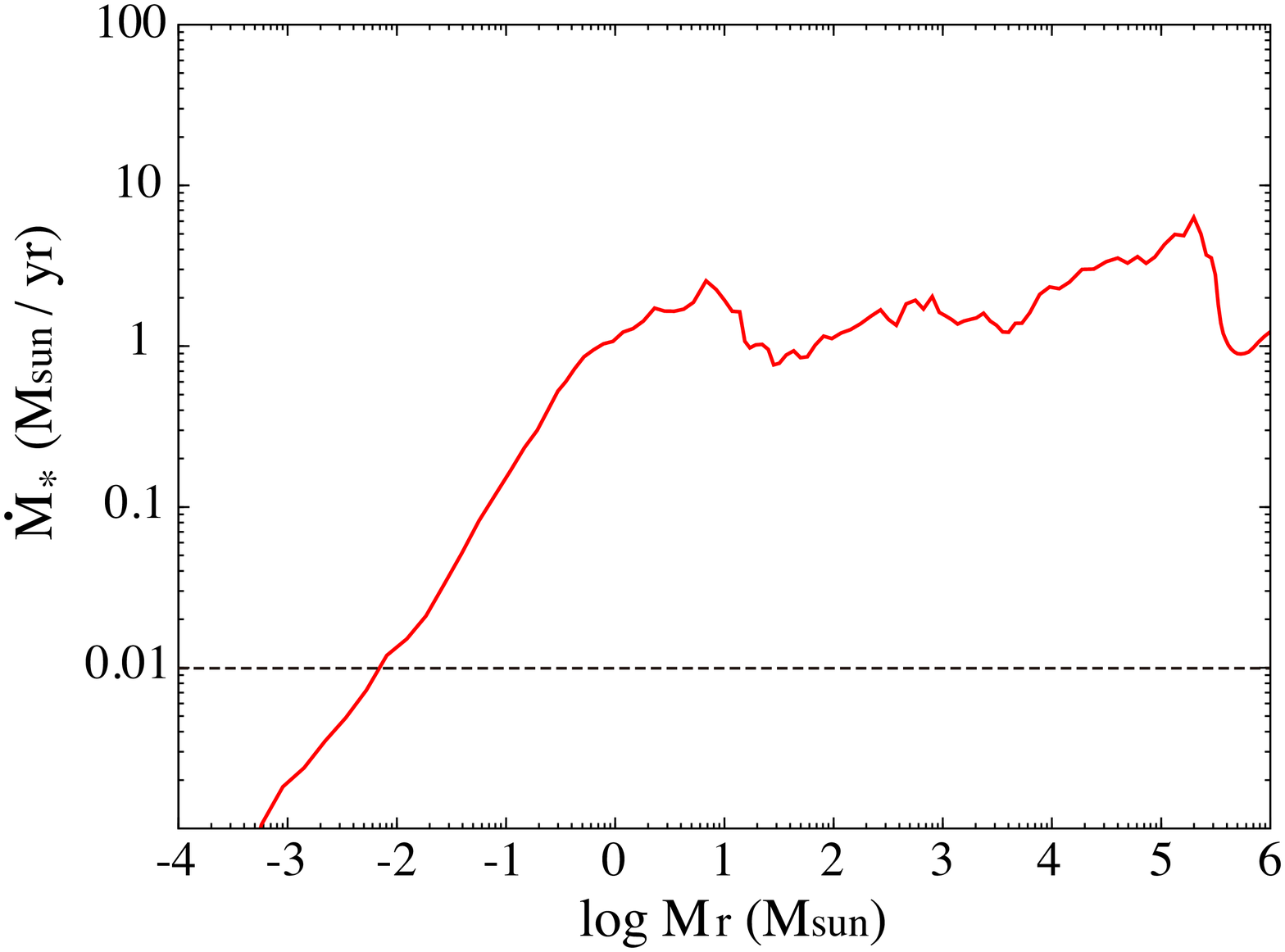}
\end{center}
\caption{Profile of the mass infall rate ($\mdot=-4\pi \rho r^2 v_{\rm rad}$)
as a function of the enclosed mass at the end of the simulation.
The horizontal line indicates the critical value $10^{-2}~\msunyr$, 
above which the protostar swells to supergiant and the radiative feedback is 
strongly suppressed. 
\citep{HOY12}.
}
\label{fig:mdot}
\end{figure}

Fig.~\ref{fig:r_vel} presents the profiles
of the radial and tangential velocities.
Also shown for comparison is half of the Keplerian velocity 
and the sound speed.
Both the radial and tangential flows become supersonic with the Mach number of $2-3$.
At the surface of the protostar, the radial flow is abruptly brought to a halt.
In the accreting envelope, the tangential velocity is as large as half the Keplerian velocity,
in accordance with previous studies of Pop III star formation
\citep[e.g.,][]{2002Sci...295...93A, 2008Sci...321..669Y}.
It is known that the cloud can contract in the runaway fashion
even with conserved angular momentum
as long as the temperature does not increase with the density
\citep[e.g.,][]{1984PThPh..72.1118N, 1998ApJ...493..342S}.
We note that the turbulent velocity is accelerated to the well-known universal value of $\sim 0.5~v_{\rm Kep}$ 
soon after the gravitational collapse starts even with initially weak turbulence.
Thus, the result seems unlikely to depend on the initial turbulent velocity.
After protostar formation, on the other hand, materials initially located 
in the outer radius and thus with higher specific angular momentum begin to fall in, 
and the centrifugal radius increases outwards
\citep{2000ApJ...531..971S}.
In our simulation, however, the rotationally supported disk has not yet appeared
because the centrifugal radius ($\la 0.1$ AU; \citealt{2008ApJ...681..771M})
is still smaller than the stellar radius $\sim 1$ AU
in this early accretion phase.

Fig.~\ref{fig:mdot} shows the radial profile of the mass infall rate
as a function of the enclosed mass.
This can be regarded as the temporal evolution 
of the accretion rate after the protostar formation.
Note that the value inside the protostar ($M_r \la 1~\msun$) 
is not equivalent to the accretion rate.
The infall rate becomes almost constant for the flat temperature profile
because in the self-similar solution, the flat temperature profile
is proportional to $c_s^3/G$,
which depends only on the temperature of the accreting envelope.
The typical value is as high as $\sim 2~\msunyr$, 
which is consistent with the previous simulations starting from the cosmological 
initial condition \citep[e.g.,][]{2013MNRAS.433.1607L}.
This infall rate is larger than $20~c_s^3/G$ for $T=8000$ K
and similar to the value found for the runaway collapse starting from 
an initial condition not so far from the hydrostatic equilibrium.
\citep{1993ApJ...416..303F}.
The protostar is expected to grow via such rapid accretion to an SMS
within its lifetime $\sim 10^6$ yr.
When the stellar mass exceeds $\sim 10^5~\msun$, the SMS is expected to
collapse to a BH by the general relativistic instability
\citep{1964ApJ...140..417C,2013ApJ...778..178H}.


\section{Conclusion and Discussion}

We simulated the collapse of a massive ($\ga 10^5~\msun$) and warm ($\sim 8000$\,K) primordial-gas cloud using idealized initial conditions with a weakly turbulent field. We found that the cloud collapses nearly isothermally through H-atomic cooling and does not undergo a major episode of fragmentation, despite the inclusion of turbulence.
Finally, a small protostar with mass $\sim 0.2~\msun$ is formed
when the central part becomes optically thick to the continuum radiation at $\rho \ga 10^{-8}$ g cm$^{-3}$ and grows to the mass $\simeq 1~\msun$ and radius $\simeq 2$ AU by the end of the simulation.

Once formed, the protostar grows via rapid accretion from the dense filamentary flows at an approximately constant rate $\sim 2~\msunyr$. Where the accretion rate is higher than $10^{-2}~\msunyr$ (dashed line in Fig.~\ref{fig:mdot}), the protostar is known to develop a giant-like structure with a bloated stellar envelope and contracting central core \citep{HOY12, 2013ApJ...778..178H} and 
grows while avoiding the significant mass-loss due to the stellar pulsation \citep{2013MNRAS.431.3036I}.
Since the effective temperature of such a supergiant protostar is $\la 10^4$ K, the UV feedback is unlikely to prevent the mass accretion on to the star. 
However, recent simulation by \cite{2014MNRAS.439.1160R} suggests the possibility of disk fragmentation around the protostar. Further simulations of the disk and the fragments with proper treatment of the H2-line cooling are needed to see whether such a high accretion rate continues to be maintained.
After the protostar grows to an SMS ($\ga 10^5~\msun$) via rapid accretion,
it eventually collapses through the general relativistic instability
to turn into a seed of high-$z$ SMBHs
\citep[e.g.,][]{2011Natur.474..616M}.

In this {\it Letter}, we started the calculation from 
the initial condition of a critical BE sphere with turbulence,
and found a single protostar formed without 
a major episode of fragmentation.
Since at $\la 0.1$ pc the profiles of the density and tangential velocity 
converge to self-similar forms with 
$\rho \propto r^{-2}$ and $v_{\rm tan}\simeq 0.5~v_{\rm Kep}$, 
independent of the initial conditions
(Figs.~\ref{fig:pro}a and \ref{fig:r_vel}),
we expect that our conclusions depend only weakly on the initial setup.
To confirm this speculation, we need to investigate the dependence on the initial conditions.
In particular, strong turbulence could prompt the efficient fragmentation,
instead of forming a single SMS (e.g., \citealt{2011ApJ...727..110C}).
Similarly, SMS formation from the proper cosmological initial condition remains to be explored for future studies.

In this simulation, we have neglected the effect of magnetic fields. 
Previous studies suggest that magnetic field strength could rival 
that of the turbulent energy 
\citep[e.g.,][]{2011ApJ...731...62F, 
T12, 2013MNRAS.432..668L}
with the effect of either increasing accretion efficiency 
(via magnetic breaking producing a more spherical flow) 
or decreasing it via protostellar jets \citep{2006ApJ...647L...1M}. 
This exploration will be left for future investigations.

\section*{Acknowledgements} 
We thank the Enzo and yt support teams, 
especially Brian O'Shea and Matthew Turk for their useful advice.
We also thank Takashi Nakamura, Takashi Hosokawa,
Naoki Yoshida, Shu-ichiro Inutsuka, Tsuyoshi Inoue, 
and Kei Tanaka for their fruitful discussions.
The results are analyzed using the visualization toolkit for astrophysical data YT
\citep{2011ApJS..192....9T}.
Numerical computations were carried out on Cray
XC30 at the Center for Computational Astrophysics
of the National Astronomical Observatory of Japan.
This work is supported in part by the grants-in-aid by the Ministry 
of Education, Culture, and Science of Japan (KI 23$\cdot $838;
KO 21684007, 25287040)

\end{document}